\documentclass[usenatbib,useAMS]{mn2e}
\usepackage{amssymb,epsfig,rotating}

\usepackage{txfonts}
\usepackage{graphicx}
\def\msun{{\rm M_{\odot}}}

\title [X-ray proplyds ]
{Probing X-ray photoevaporative winds through their interaction with
ionising radiation in cluster environments: the case for X-ray proplyds}
\author[Clarke \& Owen]{C.J. Clarke$^{1,3}$\thanks{E-mail:cclarke@ast.cam.ac.uk} and
J.E. Owen $^{2,3}$  \\
$^{1}$Institute of Astronomy, Madingley Rd, Cambridge, CB3 0HA, UK \\
${2}$ Canadian Institute for Theoretical Astrophysics, 60 St. George Street, Toronto, M5S 3H8, Canada.\\
$^{3}$ Nordita, Roslagstullsbacken 23, SE-10691, Stockholm, Sweden.}

\pagerange{\pageref{firstpage}--\pageref{lastpage}} \pubyear{2003}

\begin{document}
\def\lta{\mathrel{\spose{\lower 3pt\hbox{$\mathchar"218$}}
     \raise 2.0pt\hbox{$\mathchar"13C$}}}
\def\gta{\mathrel{\spose{\lower 3pt\hbox{$\mathchar"218$}}
     \raise 2.0pt\hbox{$\mathchar"13E$}}}
\def\Msun{{\rm M}_\odot}
\def\msun{{\rm M}_\odot}
\def\Rsun{{\rm R}_\odot}
\def\Lsun{{\rm L}_\odot}
\def\19{GRS~1915+105}
\label{firstpage}
\maketitle

\begin{abstract}
We show that if young low mass stars are subject to vigorous X-ray
driven disc winds,
then such winds may be rendered
detectable in cluster environments through their interaction with
ionising radiation from massive stars. In particular we argue that in
the ONC (Orion Nebula Cluster) one expects to see of order tens of
`X-ray proplyds'  (i.e. objects with offset ionisation fronts
detectable through optical imaging) in the range $0.3-0.6$pc from $\theta_1$C Ori (the
dominant O star in the ONC). Objects at this distance 
lie outside the central
`FUV zone' in the ONC where proplyd structures are instead well explained  by
neutral winds driven by external Far Ultraviolet (FUV) emission from
$\theta_1$C Ori.
We show that the predicted numbers and sizes of X-ray proplyds  
in this region are compatible with the numbers of proplyds observed and that
this may also provide an explanation for at least some of the far flung
proplyds observed in the Carina nebula.  We compare the sizes
of observed proplyds outside the FUV region of the ONC
with  model predictions based on the  
current observed X-ray luminosities of these sources ( bearing in mind that
the current size is actually set by the X-ray luminosity a few hundred
years previously, corresponding to the  flow time to the ionisation front).
We discuss whether variability on this timescale can plausibly explain
the proplyd size data on a case by case basis. 
We also calculate the predicted
radio free-free emission signature of X-ray proplyds and show that this
is readily detectable. 
Monitoring is however required in order to
distinguish such emission from non-thermal radio emission from active
coronae.  We also predict
that it is only at distances more than  a parsec from $\theta_1$C Ori  that the
free-free emission signature of such offset ionised structures would be
clearly distinguishable from an externally driven ionised disc wind. 
We argue that the fortuitous  proximity of massive stars in the ONC
can be used as a beacon to light up internally driven X-ray winds
and that this represents a promising avenue for observational tests of
the X-ray photoevaporation scenario. 
\end{abstract}


\section{Introduction}

The term `proplyd' was originally coined by O'Dell et al (1993) as a
contraction of `protoplanetary disc' and has been  applied  to systems
 in star formation regions imaged either  as bright rimmed cometary structures 
(in  emission lines and continuum) or  as dark `silhouette discs'. 
The
vast majority of proplyds have been detected in the Orion Nebula Cluster,
henceforth ONC (Lacques \& Vidal 1979,
Churchwell et al 1987, O' Dell et al 1993, Bally et al 2000) but
there are also examples in other regions such as Carina (Smith et al 2003),
NGC 3603 (Brandner et al 2000) and Cyg OB2 (Wright et al 2012); see also
Stecklum et al 1998, Yusef-Zadeh et al 2005,  Balog et al 2006, 
Koenig et al 2008. The common feature of such regions is the presence of 
massive (OB) stars. This provides a bright background (in the
case of silhouette discs)  and  is also consistent with the interpretation of bright cometary rims as stemming from  the interaction between ionising radiation
from these OB stars with the neutral material contained within the proplyd.
The lack of a confining medium and the fact that the escape temperature
at the ionisation front is much less than $10^4$K suggests that the gas is in
a state of expansion at that point and this has been confirmed through emission line imaging (Henney \& O' Dell 1999). The expansion timescale is so short
however that this scenario requires that neutral gas is constantly re-supplied
from some reservoir within the proplyds.

  It is likely that observed proplyds are heterogeneous in nature. For example
the `giant proplyds' (i.e. those on a scale of $10^4-10^5$ AU) in NGC 3603 and 
Cyg OB2 are likely to be irradiated clumps of molecular gas (sometimes termed
EGGs: embedded gaseous globules after Hester et al 1996). In other words, in these sources
the reservoir of neutral material is distributed throughout the proplyd volume
as would be expected if the object was pre-stellar or protostellar
in nature. This 
interpretation is consistent with the lack of stars detected within some of
these 
objects together with the rather high masses (of order a solar mass) detected
via molecular line observations (Sahai et al 2012a,b). On the other hand, the
bulk of proplyds in Orion (which are on a much smaller, $\sim 100$ AU
scale) appear to be very different in nature since they
typically contain stellar sources;  discs, where detected in silhouette 
in these systems,
are on a scale considerably less than the proplyd radius (Vicente \& Alves 2005) and are moreover very low in mass (Mann \& Williams 2010). In these objects it is necessary to posit a mechanism that lifts material from the relatively compact
disc reservoir and brings it up to the ionisation front. Johnstone et al (1998)
presented a simple and elegant framework for such objects in which 
material is lifted in a thermal wind driven by far-ultraviolet (FUV)
radiation from the massive star that also provides the source of ionising
photons. Such an interpretation is supported, for example, by the detection of
molecular hydrogen $2.1 \mu$m emission (which is known to be pumped by
FUV radiation) in a layer coincident with the disc surface (Chen et al 1998). 
Such models
provide an excellent fit to spectroscopically determined mass loss  rates
in proplyds (Richling \& Yorke 2000)
 and also broadly account for the observed proplyd size
distribution.  A critical aspect of such models is that FUV driven proplyds 
should only arise within a central zone of the cluster where the FUV radiation
field is strong enough to drive a significant neutral wind. Outside
this central `FUV zone', molecular self-shielding limits the density of the
neutral disc wind so that it can be readily penetrated by ionising photons.
In this case, the ionisation front is expected to be virtually coincident with 
the disc surface rather than being spatially offset as in many observed proplyds.
St\"orzer \& Hollenbach (1999) estimated the radius of the FUV zone
in Orion as being 0.3 pc and this is indeed consistent with the observed
concentration of proplyds within this region (Bally et al 2000).
However, there are at least $\sim 10$ proplyds in Orion that are well beyond
the FUV zone of the most massive star ($\theta_1 $C Ori) even in projection
and which are likewise too far from other massive stars in the region
to be good candidates for FUV driven winds (Vicente \& Alves 2005). 
Smith et al (2003) similarly drew attention to a population of larger proplyds
(size $\sim 900-2500$ A.U.)
in Carina that are at surprisingly large distances (up to 40 pc) from
the cluster core (note however that at least some of these may - unlike
the Orion proplyds - fall into the EGG category described above: see
Sahai et al 2012b for a demonstration of a high molecular gas mass in one
such object).

 Here we propose an alternative origin for some of the proplyds that are
observed outside the central FUV zone of star-forming regions. We point
out that if discs are subject to vigorous mass loss from winds
driven by X-rays from their central stars (as proposed by Owen et al 2010,2011,
2012)
then the interaction between such winds and the ionising radiation
from the dominant O star in the region should give rise to proplyd-like
structures. We emphasise that the mass loss rates  of X-ray driven winds
are independent of the position in the cluster and depend only
on the (time-averaged) X-ray luminosity of each source. The role of the
OB stars in the centre of the cluster is simply to `light up' the
surface of the wind at a distance that is - in the case of the more
luminous X-ray sources - resolvable in HST images of Orion. In section
2 we discuss the X-ray properties of stars in the ONC and their variability
and in Section 3 we assess the expected population of `X-ray proplyds'
and compare with observations.  In Section 4 we consider the possible
signatures of X-ray proplyds in the thermal radio continuum.
Section 5 summarises our conclusions.

\section{ The X-ray properties of stars in the Orion Nebula Cluster}

 Before embarking on an analysis of how X-ray driven winds would
affect the nature of objects imaged in the ONC we must consider
the available data on the X-ray emission of young stars in the region.
This has been well characterised by the Chandra Orion
Ultradeep Project (COUP): Preibisch et al (2005) contains an
analysis of the nearly $600$ sources  in Orion that are associated with
stars that are well characterised in the optical and among which the
X-ray detection rate is high ($> 97 \%$). Kastner et al (2005) presented
COUP data on the population of objects in Orion that are classified
as proplyds (i.e. through association with cometary rims and/or
silhouette structures). The detection rate in this sample (which overlaps
the optical sample described above and which comprises $\sim 140$ objects)
is considerably lower ($\sim 70 \%$):  the correlation between
$N_H$ and inclination in silhouette discs (noted by  Kastner et al 2005)
suggests that this is due to absorption rather than necessarily
implying that proplyds are intrinsically weaker in the X-ray.
 
 For clarity, we here stress that we do {\it not} define an X-ray proplyd
as being a proplyd with a detected X-ray flux. Throughout this paper
we use the term  X-ray proplyd to denote {\it an object with an offset ionisation
front that can be attributed to the interaction between ionising radiation
from OB stars and a neutral wind driven off the disc by X-ray photoevaporation}
(see Figure 1).

  Before assessing whether there is any evidence for such a population
in Orion we need to consider the amplitude  of X-ray variability. When
we come to calculate the expected radii of offset ionisation fronts
as a function of X-ray luminosity (see equation (3), Section 3), 
the value of $L_X$
that enters this calculation represents the X-ray luminosity 
(averaged over the thermal timescale at the flow base) 
 that was emitted at a previous epoch separated in time
by the timecsale for the X-ray wind to propagate out to the
ionisation front. This flow time is typically a few hundred years. 
We obviously have no
information on the level of X-ray variability on these timescales. 
Although Favata et al (2004) found that some objects undergo
variations of an order of magnitude or more on a timescale
of years, Micela \& Marino (2003) reported that the level of variability over
periods from 
months to years is typically around $3-4$   while 
Preibisch et al (2005) found a median change in X-ray luminosity of
around a factor two between the COUP observations and those
of Feigelson et al 2002 obtained $4$ years earlier. The most recent
variability study in Orion (Principe et al 2014) finds that variations
by an order of magnitude or more are common over a four year period. 
 Preibisch
et al noted  the large scatter
in X-ray luminosities as a function of all variables (such as stellar mass,
rotation period or Rossby number) and pointed out  that this could
in principle be due to large amplitude variability cycles.  The
magnitude of this scatter (i.e. more than two orders of magnitude in $L_X$
at
given stellar mass) however means that one would need to invoke 
large amplitude variability
over a longer timesccale than that covered by observational studies.

 Given the lack of observational constraints on the level of
X-ray variability over hundreds of years,  we need
to quantify the possible role of variability indirectly. In
Section 3.1  we present  a statistical approach (which assumes
only that the X-ray luminosity function of the entire population
is invariant over hundreds of years); in Section 3.2 we examine
individual objects on a case by case basis,
assessing the level of variability that is required in order
for the observations and model predictions to agree.

\section{The size of X-ray proplyds as a function of X-ray luminosity and
distance to the ionising star}

  We consider the situation where a star of X-ray luminosity $L_X$
is located at a distance $d$ from an ionising source  with ionising 
photon output rate of $\Phi_{ion}$ s$^{-1}$.
Following Owen et al (2012) we write

\begin{equation}
\dot M_w = 8 \times 10^{-9} L_{X30} M_\odot {\rm{yr}}^{-1}
\end{equation}

\noindent (where $L_{X30}$ is the X-ray luminosity in units of
$10^{30}$ erg s$^{-1}$) noting that the mass loss rate is independent of the stellar mass 
and also of disc mass and radius, provided the latter is greater than $30-40$ A.U.
for a solar mass star (see Figure 4 of Owen et al 2010 for a cumulative
mass loss profile from the disc, bearing in mind that the radial
coordinate in this figure scales linearly with the stellar mass). The relevant
X-ray luminosity 
 is its  value (averaged over
a thermal timescale at the flow base: less than  a decade, Owen et al 2010) at the time in the past ( of order a few hundred years ago) when material 
currently on a $\sim 100$ A.U. scale was launched from the disc.

   This X-ray driven wind is largely neutral but at some radius, $R_{IF}$, it
becomes ionised by 
radiation from the dominant OB star
in the cluster. Outward of $R_{IF}$ the flow expands in a spherical transonic
flow (i.e. with velocity of order the sound speed in the ionised medium).
The ionisation front constitutes a contact discontinuity of the flow,
requiring
that the neutral flow entering the ionisation front is sub-sonic with
respect to the neutral gas. In the absence of external ionisation, the
X-ray driven wind becomes supersonic at radii less than $50$ A.U. (i.e.
on scales that are unresolvable even in the case of HST observations
of the Orion Nebula Cluster). 
Observable proplyds must therefore correspond to the case where an 
already supersonic neutral flow has to undergo a shock in order to
deliver neutral gas to the ionisation front with subsonic velocities. This
is a qualitatively identical situation to that described by Johnstone
et al (1998) in the case of ionising radiation interacting with
FUV driven winds: clearly the details of the shock location will
differ since the X-ray driven wind is considerably warmer ($\sim 4000$ K)
than that
in the FUV driven flows ($\sim 1000$ K).
We show a schematic diagram of the flow in an X-ray proplyd in Figure 1.

\begin{figure}
\includegraphics[width=8.cm]{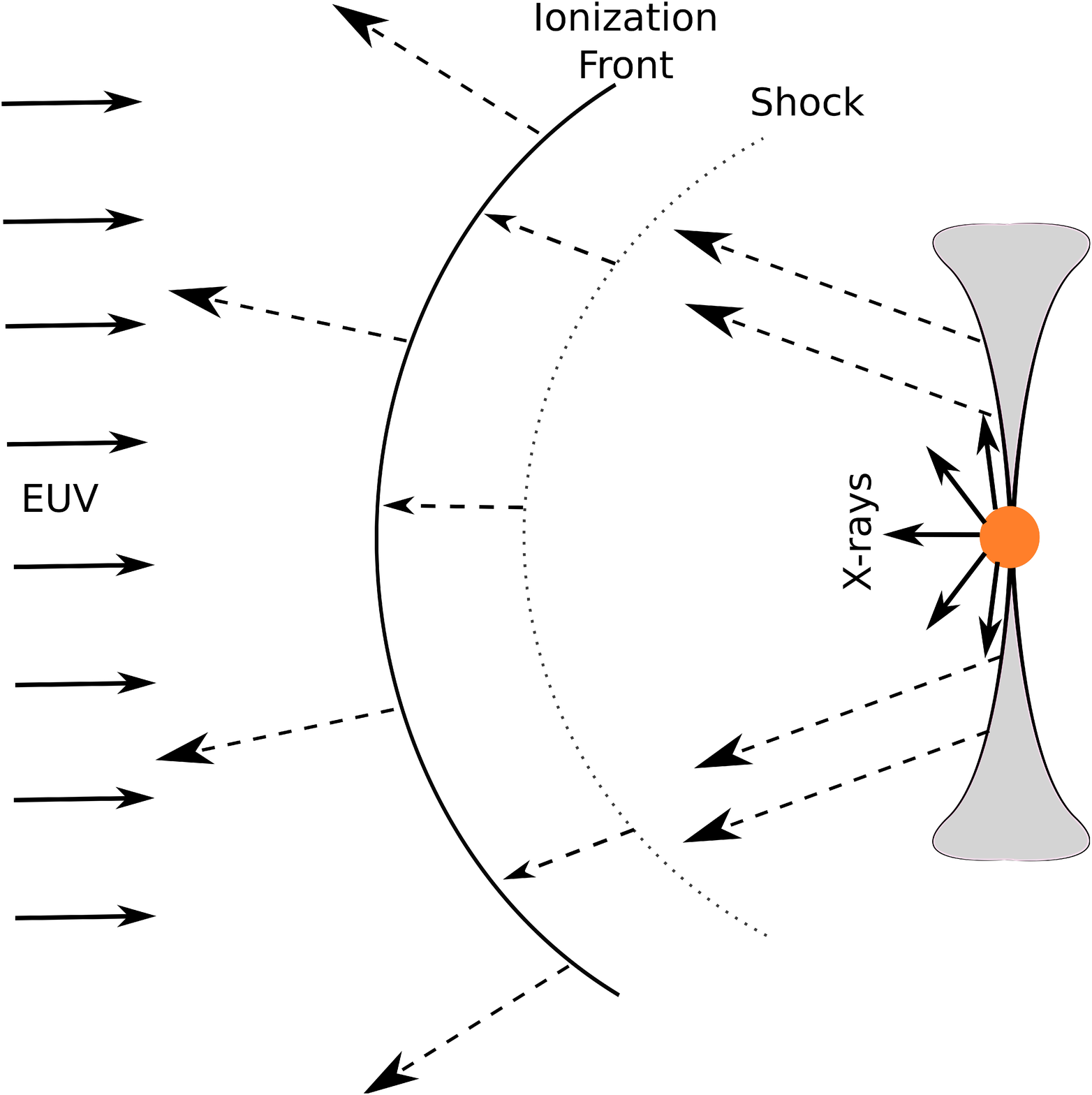}
\caption{Schematic of flow in the case of an X-ray driven wind from
a disc interacting
with the ionising (EUV) radiation from a massive star.}
\end{figure}   
  What concerns us here is however the location of the ionisation front
which is set by the requirement that the integrated recombination rate in
the ionised flow matches the input of ionising photons (see Churchwell et
al 1987 for a demonstration that the fraction of ionised photons that
are additionally required to ionise the flux of neutral material into
the ionisation front is a small fraction of the above). Since the
density in the spherical transonic ionised wind falls with radius as
$r^{-2}$,  the integrated recombination rate per unit area scales
as $n_i^2 R_{IF}$, where $n_i$ is the density of ionised gas at
$R_{IF}$ (i.e. it is dominated by conditions close to the
ionisation front). This condition fixes $n_i$ as a function of
$R_{IF}$ and ionising flux. The mass loss rate from a spherical
transonic ionised flow is however given by

\begin{equation}
\dot M = 4 \pi R_{IF}^2 n_i \mu m_H c_s
\end{equation}

\noindent where $\mu$ is the mean molecular weight, $m_H$ is the mass of a proton
and $c_s$ is the sound speed in the ionised gas. Combined with the
ionisation equilibrium requirement above, this fixes $R_{IF}$ as a function
of $\dot M$ {\it independent of the mechanism that delivers material
to the ionisation front}. We can thus readily adapt the formulation
of Johnstone et al (1998) (where the FUV mass loss rate is determined
by the requirement of a fixed neutral column) to the case of X-ray
driven mass loss, which depends only on the X-ray luminosity
of the source (see equation (1)). We thus have that the `chord diameter'
of the proplyd (which is equal to $2.6 \times R_{IF}$; Vicente \& Alves
2005) is given by:

\begin{equation}
R_{cd} =  225 {\rm{A.U.}} \biggl({{d_{pc}^2 L_{X30}^2}\over{\Phi_{ion49}}}\biggr)^{1/3}
\end{equation}

\noindent where $d_{pc}$ is the distance from the ionising source in parsecs
and 
$\Phi_{ion49}$ is the ionisation rate expressed in units of 
 $10^{49}$ s$^{-1}$.

 If we now apply this estimate to Orion (where $\Phi_{ion49}$ is about
1 in the case of the dominant OB star in the Trapezium cluster,
$\theta_1 $C Ori), we see that a proplyd with 
$L_{X30} \sim 1$ (which lies towards the upper end of the observed X-ray
luminosity function for accreting stars in Orion; Preibisch et al 2005) would
give rise to a proplyd of chord diameter $ \sim 100$ A.U. if placed
at the outer edge of the `FUV zone' (i.e. at $0.3$ pc from
$\theta _1$C Ori). Since the cumulative X-ray luminosity function
rises very steeply towards higher luminosities, it is unlikely that
many of the observed proplyds in the FUV zone
could instead be powered by internal X-ray driven winds 
( around $80 \%$ of stars observed by HST in this region are found to have proplyds
that are considerably larger than this, i.e. with chord diameters of
many hundreds of A.U.). On the other hand, models for proplyds
in this region where disc winds are driven by FUV radiation from
$\theta_1$C Ori can well account for the sizes of the observed proplyd
distribution (Johnstone et al 1998). We thus concur with previous authors
that the origin of the proplyd population within $0.3$ pc of 
$\theta_1$C Ori is indeed likely to be a disc wind driven by external
FUV radiation.

 Although the majority of proplyds in the ONC are located in this
central FUV zone, there are a number of proplyds at larger radius,
even in projection. Around half of these may still be explicable
by FUV heating by other OB stars in the region (i.e. $\theta_1$A Ori and
$\theta_1$B Ori) although this has not been demonstrated. There are
however at least $9$ sources that are located in the range $0.3-0.6$ pc
{\it in projection} from $\theta_1$C Ori and which are likewise
unlikely to receive significant FUV heating from the other OB stars.
In Section 3.1 we assess whether the number and sizes of such objects
are consistent with the expectations of X-ray driven proplyds, assuming
that the X-ray luminosity function a few hundred years ago (when the current proplyd size was
set) is identical to what it is now. This is a minimally constraining
plausible assumption and so satisfying this test is a necessary
condition for the viability of the model.
In Section 3.2  we compare the observed X-ray fluxes of these $9$ sources
with the values that are required to explain their proplyd sizes and
discuss whether any differences can be plausibly ascribed to variability.

\subsection{Monte Carlo simulation of X-ray proplyds in the Orion Nebula
Cluster}


In what follows we adopt equation (3) in order to determine the expected
chord diameter of a proplyd of given $L_X$ and distance $d$ from
$\theta_1 $C Ori. We populate stars in a spherically symmetric distribution
centred on $\theta_1$C Ori  which is consistent with the radial dependence
of the observed surface density profile (Jones \& Walker 1988,
Hillenbrand 1997). We need to ensure that the density normalisation is
consistent with the sample of objects contained in HST imaging
campaigns of the ONC. Vicente \& Alves record around $100$ proplyds
imaged in the central $0.3$ pc of the cluster and comment that
this is around $80 \%$ of the stars observed by HST in this region. Our
piecewise power-law density distribution is thus normalised
so that there are $\sim 120$ stars in this region:

\begin{eqnarray}
\rho_*=500 {\rm{pc}}^{-3} \biggl({{0.3 {\rm{pc}}}\over{d}}\biggr)\\
\rho_*=500 {\rm{pc}}^{-3} \biggl({{0.3 {\rm{pc}}}\over{d}}\biggr)^{2} \\
\rho_*= 45 {\rm{pc}}^{-3} \biggl({ {{\rm{pc}}}\over{d}}\biggr)^{5}
\end{eqnarray}

\noindent for the three regimes $d < 0.3$pc, $0.3 < d < 1$ pc and $d>1$ pc respectively;
this roughly mimics the observed projected source distribution of Jones
\& Walker 1988 (see e.g. Figures 3 and 4 of Scally et al 2005).
In our population synthesis modeling described below we also require
that X-ray proplyds are only found in systems with discs and so
multiply the above density profile by the observed disc fraction, $f_d$:
we adopt $f_d=0.8$ for $d< 0.3$ pc and $f_d = 0.7$ for $d > 0.3$ pc
(Hillenbrand \& Hartmann  1998).

  For each star we select an X-ray luminosity from the COUP
X-ray luminosity function for the ONC (Preibisch et al 2005); 
since X-ray proplyds can only
be produced by stars with discs we adopt a  parameterisation
of the luminosity function for `accretors' shown in Figure 17
of Preibisch et al 2005,  limiting ourselves to stars with $L < 5 L_\odot$,
corresponding to stars less massive than $\sim 2 M_\odot$.  (Note that we do not use the XLF for proplyds
partly because if the relative incompleteness of this sample in the X-rays -
see discussion in Section 2 - and also because we need for this exercise
to use the XLF of the `parent' disc bearing population rather than
those that have been selected on account of resolvable structures). We
parameterise the accretor XLF in three sections that are each individually
flat in log$(L_X)$, corresponding to the intervals $28-30$, $30-30.3$  and $30.3-31$.
The cumulative fractions of sources with log $(L_X)  < 30.$, $<30.3$   and $< 31.$ are respectively
$0.85$, $0.96$ and $1$. Given that equation (3) implies that resolvable proplyds
are associated with objects with $L_X > 10^{30}$, this analysis is evidently only
sensitive to the parameterisation of the upper regions of the cumulative
distribution function. For this exercise (in which we are 
estimating total numbers of predicted resolvable proplyds) we do not further
sub-divide the 
Preibisch data by mass though we note that
the upper envelope of the XLF declines rather
steeply for spectral types of M4 and later. We will bring this consideration to bear when assessing the
individual sources in Section 3.2 below. 

 We then evaluate the expected
chord diameter based on equation (3). According to Vicente \& Alves (2005),
the proplyd census is complete only for structures larger than $150$ A.U..
We thus record the projected position and chord diameters of all proplyds
that are larger than this.

  The result of this exercise is that we predict  $\sim 20-30 $ X-ray proplyds larger than $150$ A.U. at projected distances of $0.3-0.6$pc from  $\theta_1$C Ori; within
$0.3$ pc we predict $\sim 10$ such proplyds. (As noted above, the latter is very small
compared with the large number of proplyds ($\sim 100$) observed in this
central region
and confirms the conventional interpretation of such objects as
being driven by external FUV photoevaporation). The
expected number of X-ray proplyds {\it outside} the central FUV zone
is consistent with the numbers observed. The comparison with
observations is complicated by the fact that it is unclear exactly
how many of the proplyds observed outside the FUV zone of $\theta_1$C Ori  can
instead be attributed to FUV heating by other OB stars in the cluster. In
Figure 2 we present the observed size distribution of proplyds that
are possible candidate X-ray proplyds under two assumptions a) that all
proplyds ($\sim 35$) more than $0.3$ pc from $\theta_1$C Ori  are in
this category and b) instead omitting from this sample those proplyds that
are in the SW quadrant (i.e. in the vicinity of $\theta_1$A Ori and
$\theta_1$B Ori). Under this conservative assumption $\sim 9$
proplyds are candidate X-ray proplyds (see Table 1) and these
are described by the bold histogram in Figure 2.
We see that the predicted numbers and size distributions of  X-ray
proplyds are well bracketed by the observations.

  The Monte Carlo analysis has therefore demonstrated that the numbers
and sizes of proplyds observed in this region are broadly consistent with those
predicted by the X-ray proplyd model.
\subsection{Assessing individual candidate X-ray proplyds.}

   Given this success in terms of predicting over-all numbers of
proplyds with projected separations in the range $0.3-0.6$ pc 
we now proceed to examining individual sources. We will investigate whether
the sizes of observed proplyds in this region are consistent with those
predicted from their current X-ray luminosities and, if not, what
level of variability would need to be invoked in order to bring the
figures into agreement. 
 
 As explained in Section 2, we first assume that the chord diameter
of offset ionisation fronts can be calculated from equation (3), with
$L_X$ given by the {\it observed} X-ray luminosity of each source.
In Table 1 we detail $9$ proplyds that lie outside the
central FUV zone of $\theta_1$C Ori  (i.e. outside a projected radius of
$0.3$ pc and which are not in the SW quadrant where they may be
ionised instead by $\theta_1 $A Ori  and $\theta_1 $B Ori). In $8/9$ of the sources
there is an optical counterpart and we list, where available,  the spectral type as given
by Getman et al (2005).  Observed chord diameters are obtained from Vicente \& Alves (2005). $7/9$ of the sources are detected in X-rays and for these
sources we list $L_X$ from
Kastner et al (2005) and  the minimum predicted chord diameter ($R_{pm}$)
for the X-ray proplyd model, calculated using  equation (3).
The value of $R_{pm}$ is   somewhat  under-estimated  
 since we  set the distance from $\theta_1$C Ori  equal to
the projected distance: on average this will cause the predicted
sizes to be under-estimated by a factor of order unity given
the $d^{2/3}$ scaling of equation (3). 
 Note that, in contrast, the
Monte Carlo analysis of Section 3.1 explicitly models how projection
effects influence the expected sizes distribution but that this
approach is not applicable on an object by object basis.

  We also list for each source the value of log $ L_{X,m}$ which
is the value of 
 X-ray luminosity required to bring the predicted and observed proplyd
sizes into agreement in the case that the 3D distance of the source
from $\theta_1 $C Ori  is equal to its projected distance. The final
column lists the ratio of $ L_{X,m}$ to the bolometric luminosity
of the source which is derived, where available, from Getman et al 2005.
The actual
value of the luminosity required is actually $L_{X,m}/\tilde r$
where $\tilde r$ is the
uknown ratio of the 3D distance to $\theta_1 $C Ori  and its projected value.
For the assumed cluster density profile, the expectation value of $\tilde r$ is about $1.4$ and its
maximum value is $\sim 3$. 

 There is one source (172-135) where
the  non-detection in X-rays can be explained by it being 
a silhouette disc viewed close to edge on: the
poor constraint on the intrinsic X-ray luminosity makes it unsuitable for
our analysis. Among the remaining $8$ sources, there are two where the
observed X-ray luminosity is consistent with that required to explain
the proplyd size (invoking at most modest projection corrections): these
two sources are 073-227 and 140-1952. There are three sources 
(005-514,102-021 and 152-738) where the combination of projection effects
and less than order of magnitude variations can reconcile the model
predictions with observations. The final three objects (066-652,
131-046  and 097-125) all require more than order of magnitude enhancement of the
X-day flux a few hundred years ago compared to its current value. 
Note that the
fact  that
we would require all the sources to have been as bright or brighter
in the X-ray than they are now is not {\it per se} an argument against the model: 
we would expect resolvable proplyds to be objects that 
selectively populate the upper end of the XLF at the relevant epoch and
thus, on average, that they should be fainter now than previously.  On the other
hand, we can ask whether the  X-ray luminosity that we need to
invoke in the past is a reasonable value given  
typical values of $L_X/L_{bol}$ in young
stars as a function of spectral type. The final column shows that the
values required are of order $10^{-3}$; such values are found in around
$10 \%$ of young stars  in the `accretor' category, regardless of spectral type (Preibisch et al 2005).
 
 We therefore conclude that at least some of the sample are compatible with
being X-ray proplyds:   the number that fall into this category depends
on the amplitude of X-ray variability that one is
prepared to invoke on timescales of hundreds of years.
It has to remain a matter of speculation whether individual sources
may (as in the most extreme case, 066-652) be as much as a factor
hundred  fainter currently than they were in the past; we however
note that even here the implied $L_X/L_{bol}$ value at a previous epoch
was not extreme. When considering the likelihood that these objects are
indeed X-ray proplyds it needs to be borne in mind that no other models
have been proposed to explain offset ionisation fronts at such large
distances from $\theta_1 $C Ori.

 Since we have shown in Section 3.1 that the model predicts
a roughly correct number of resolvable X-ray proplyds over-all
 (while  Table 1 implies that most of the observed proplyds need
to have been more X-ray luminous in the past than they are now) 
it is a necessary corollary that there should be objects that
are currently X-ray bright but which do not show resolvable proplyd structure.
Examples of such objects are found among  
 pure
silhouette discs 
(i.e. those that do {\it not}
show a resolvable offset ionisation front) since in some of these 
the  current X-ray luminosities  are sufficient to produce a detectable
offset ionisation front (e.g. $183-405$) {\footnote{  There may however be 
a detectability issue  in edge-on silhouettes which does not affect conventional (FUV driven)
proplyds. While in the FUV case the wind is launched from regions
extending to the disc's outer edge 
 and the ionisation front
is always somewhat larger than the disc, optical depth effects in the
X-ray driven case  restrict the wind launching region 
to $<  50$ A.U. {\it regardless of the disc size}. Due to disc flaring,
it may be difficult to detect compact 
 ($\sim 100$ A.U.) scale ionisation
fronts in the case of large, edge-on silhouettes (as in $114-426$ and $053-717$).}
 Although we need to argue that X-ray variability disrupts any strong  
correlation between X-ray luminosity and resolvable proplyd
structure,  there is  some weak evidence for a residual association:
 among our candidate X-ray proplyds
(i.e. the objects listed in Table 1), $7/9$ are detected in X-rays
whereas among the pure silhouette discs this figure is only $8/16$ (Kastner et al 2005).

\begin{table}
\centering
 \caption{ Predicted and observed properties of candidate X-ray proplyds.
ST is the spectral type, $R_{cd}$ is the chord diameter in A.U. and $R_{pm}$ is a lower limit to
the predicted chord diameter from the model (equation (3))
under the assumption that
the 3D distance to $\theta_1 $C Ori is the projected distance and the
relevant X-ray luminosity is the observed value (listed in the successive
column). $ L_{X,m}$ (next column) is the X-ray luminosity that is required in
order for the model to match the data in the case that the
3D distance to $\theta_1 $C Ori  is the projected value; the actual
predicted X-ray luminosity required is typically a few tenths of a dex
smaller than this on account of projection effects (see text).
 The final column ($\tilde L)$ lists the logarithmic ratio of $ L_{X,m}$ to the bolometric
luminosity of the source.  
$^{\it{a}}$ The non-detection of this
source in the X-ray is explicable in terms of it being a nearly edge-on
silhouette disc.}  
\begin{tabular}{rllllll}
\hline
\hline
${\rm Name }$ & ${\rm ST}$ & $ R_{cd}$ & $R_{pm} $ & ${\rm log}(L_X)$ & $ {\rm log} ( L_{X,m})$ & $ {\rm log} ( \tilde L ) $  \\
\hline
$005-514$&${\rm K}6$&$414$&$137$&$29.9$&$30.6$&$-2.5$ \\
$066-652$&${\rm M}4.5$&$612$&$34$&$29.0$&$30.9$&$-2.3$ \\
$072-135^{\it{a}}$&$-$&$472$&$-$&$-$&$30.8$&$-$ \\
$073-227$&${\rm M}2-{\rm M}4$&$207$&$142$&$30.1$&$30.3$&$-2.7$ \\
$097-125$&${\rm M}3.5$&$207$&$32$&$29.1$&$30.3$&$-3.0$ \\
$102-021$&${\rm M}3.5$&$155$&$39$&$29.2$&$30.1$&$-2.7$ \\
$131-046$&$-$&$270$&$-$&$-$&$30.5$&$-$ \\
$140-1952$&${\rm late}{\rm G}$&$228$&$219$&$30.3$&$30.3$&$-3.0$ \\
$152-738$&$-$&$243$&$73$&$29.5$&$30.3$&$-$ \\
\hline
\end{tabular}
\end{table}

\begin{figure}
\includegraphics[width=8.cm]{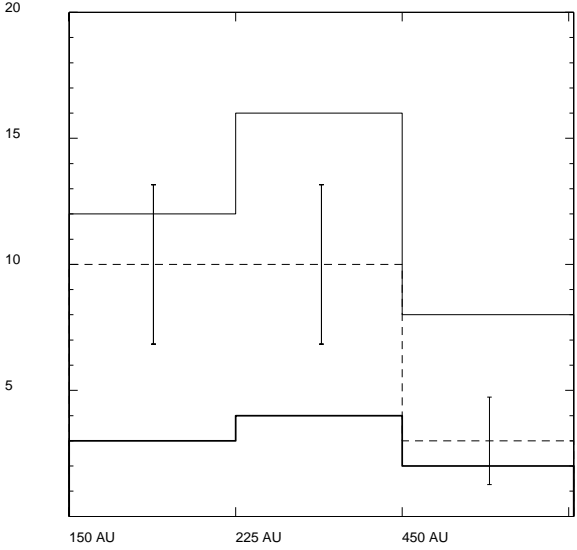}
\caption{The numbers of proplyds as a function of size in the ONC. The
upper histogram refers to the $35$ proplyds in the ONC that are 
in the range $0.3$ to $0.6$ pc from $\theta_1$C Ori. The lowest (bold) histogram
considers the same sample but omitting the $24$ objects in this
category that may arguably be in the FUV zone of $\theta_1$A Ori or $\theta_1$B Ori.
The dashed histogram is the result of the population synthesis exercise
for the predicted numbers of X-ray proplyds in the radial range 
$0.3$ to $0.6$ pc in projection. The three size bins refer to chord
diameters in the range 
$150-225$ A.U., $225-450$ A.U. and $> 450$ A.U.. Observational data
from the compilation of Vicente \& Alves (2005).} 
\end{figure}

\section{The radio emission signature of X-ray proplyds}

  We now  consider the free-free emission signature 
that we expect to be associated with the externally ionised X-ray
proplyd population ({see Pascucci et al 2012,2014, Owen et al 2013
 for a discussion of
the free-free emission signatures expected from photoevaporative
flows in the absence of external ionisation). For now we  only consider the emission from ionised
material in the wind external to $R_{IF}$. For reference we note
that X-ray driven photoevaporative flows also generate an intrinsic
free-free flux within an A.U.  associated with  ionisation by EUV photons from the star; scaling the
results of Owen et al (2013) to the distance of Orion and assuming a ratio of ionising photon output to X-ray luminosity of $10^{11}$ erg$^{-1}$, we expect the
{\it intrinsic} flux density  at $2$ cm to be:

\begin{equation}
L_{ff; int} = 0.005  L_{X30} {\rm{mJy}}
\end{equation}

 This value is comfortably less than the levels arising from external ionisation 
(equations (10) and (11) below) and so we do not consider free-free emission from internal ionisation further. 
 In order to calculate the expected free-free flux from {\it external}
ionisation of X-ray driven winds, we employ equation (2) of Garay et al
(1987) for optically thin thermal emission  {\footnote{ Note that the optical depth is proportional to the
integral of $n^2$ along the line of sight which, in ionisation
equilibrium, depends only on the ionising flux, independent of $n$;
we thus find that the predicted emission is {\it optically thin}
for wavelengths $< 15~{\rm cm} \Phi_{49}^{-1/2.1}d_{pc}^{2/2.1}$,
independent of X-ray luminosity.}}:

\begin{equation}
\biggl[{{S_\nu}\over{{\rm{mJy}}}}\biggl] = 3.4 \biggl[{{\nu}\over{{\rm{GHz}}}}\biggr]^{-0.1} \biggl[{{T_e}\over{10^4 {\rm{K}}}}\biggr]^{-0.35} \biggl[{{{\rm{VEM}}}\over{10^{57} {\rm{cm}}^{-3}}}\biggr] \biggl[{{D}\over{{\rm{kpc}}}}\biggr]^{-2}
\end{equation}
 
where $D$ is the distance to the ONC and the volume emission measure for
a spherical constant velocity wind is given by:

\begin{equation} 
VEM = 4 \pi n_i^2 R_{IF}^3
\end{equation}

 Combining this also with equations (2) and (3) (recalling that
$R_{cd} = 2.6 R_{IF}$ and assuming $T_e = 10^4$K and $D=450$pc) 
we then obtain that the flux density at $15$ GHz ($2$ cm) is:

\begin{equation}
\biggl[{{S_{15 {\rm{GHz}}}}\over{{\rm{mJy}}}}\biggr] = 0.06 L_{X30}^{4/3} d_{pc}^{-2/3}  
\end{equation}

  In Figure 3 we convert this expression into the equivalent radio luminosity
($L_{ff}$ in erg s$^{-1} {\rm{Hz}}^{-1}$) and plot the predicted relationship between
$L_{ff}$ and $L_X$ for a range of values of $d$  (neglecting X-ray variability:
see Section 3.2).
We however need to take account of the fact that - even in 
the absence of any significant neutral (X-ray driven) wind -  objects
outside the central FUV zone are expected to sustain ionised winds
associated with direct evaporation by EUV (ionising) photons
from $\theta_1$C Ori: naturally such winds do not produce a proplyd signature
(because the ionisation front is coincident with the disc surface)
but there is an associated
free-free emission signature. 
Adopting the mass
loss rates for such winds as predicted by Johnstone et al (1998) we find:
\begin{equation}
\biggl[{{S_{15 {\rm{GHz}}}\over{{\rm{mJy}}}}}\biggr]|_{EUV} = 0.06 R_{d100}^2 d_{pc}^{-2}
\end{equation}

  where $R_{d100}$ is the disc radius normalised to $100$ A.U. (note that
in this case the wind is driven over the entire extent of the disc and hence
the VEM depends on the disc radius, in contrast to the case of X-ray driven
winds: see footnote 1) 
The free-free emission levels predicted by equation (10) thus only apply
where these exceed the `floor' EUV level predicted by equation (11). 
In Figure 3 we mark on each track the floor levels of radio emission corresponding
to disc radii of $50$ and $100$ A.U. by open and filled triangles
respectively.

 We also plot (open circles) the
G\"udel \& Benz (1993) relationship between X-ray and radio
luminosity  based on
observations of weak line T Tauri stars and other disc-less but
active objects. This characteristic locus is generally interpreted in
terms of the generation of both X-ray and (non-thermal) radio emission
in the stellar corona. The solid dots and open squares
are derived from the VLA survey of the innermost region 
of the ONC by Zapata et al (2004), with the latter
symbols corresponding to optical proplyds where the X-ray
luminosities are taken from Kastner et al (2005). We emphasise that
this data cannot be contrasted with our model predictions because all the  
sources in the VLA survey have projected distances that place
them within  the central FUV zone. The data however demonstrates that
the bulk of non-proplyd sources lie along the G\"udel-Benz relation
(as expected in the case of disc-less stars) whereas most
optical proplyds are indeed
well offset to the right of this 
relation ( see Forbrich \& Wolk 2013), consistent with there being
an additional
thermal component as  argued by Churchwell et al 1987,
Garay et al 1988. The free-free flux density of the brightest proplyds
in the central FUV zone of the ONC is several $10$s of mJy and thus much
greater than what we are predicting in the case of X-ray proplyds.
(In the case of a transonic spherical ionised wind, the radio luminosity
scales with the square of the mass loss rate and inversely with the radius
of the ionisation front: the proplyds in the `FUV zone' have high
mass loss rates and are relatively compact and are thus considerably
brighter in the radio than we predict for the putative population of
X-ray proplyds at larger radius in the cluster). 

\begin{figure}
\includegraphics[width=8.truecm]{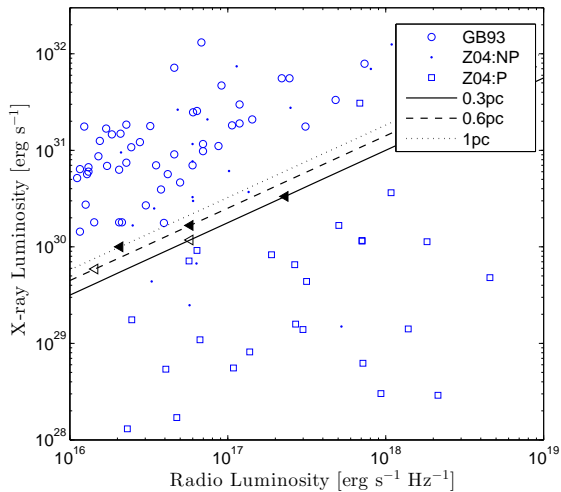}
 \caption{The predicted relationship between $3.6$ cm radio luminosity
(in erg s$^{-1}$ Hz $^{-1}$) and X-ray luminosity (in erg s $^{-1}$)
for a range of distances from $\theta_1 $C Ori. The open and filled
triangles  on each track correspond to the `floor' values of
radio emission (for disc radii of $50$ and $100$ A.U. respectively)
that are expected from ionised winds driven by 
 $\theta_1$C Ori in the absence of internal X-ray driven winds.
The open circles  are
the data of  G\"udel \& Benz (1993), while the remaining points
are from the survey of Zapata et al (2004) for the inner regions (`FUV
zone') of the ONC, with  squares and  solid  dots 
denoting sources
that are (Z04:P) or are not (Z04:NP) identified with imaged proplyd systems.
X-ray luminosities for the proplyd systems derive from Kastner et al (2005)}
\end{figure}

The predicted level of thermal radio emission from
X-ray proplyds is in the range $0.1-1$ mJy and would be readily
detectable in the ONC.  
 There are however two factors  (confusion with non-thermal emission
in X-ray luminous stars and also confusion with emission
from ionised winds driven by $\theta_1$C Ori) that complicate
the  interpretation of sources detected
along the predicted loci.
 Since our predicted thermal emission signature (equation (10)) scales as  $ L_X^{4/3}$ 
(at fixed distance from $\theta_1$C Ori), it     defines a relation that
is roughly parallel to the G\"udel \& Benz  relation. At distances
$> 0.3$ pc from $\theta_1C$ Ori  (which is the region in which
we expect X-ray proplyds to take over from FUV driven proplyds) the
expected relation is close to the bottom of  the G\"udel \&
Benz relation. We thus conclude that the thermal
emission from such objects would be hard to distinguish from
non-thermal (coronal) emission based on radio
luminosities alone.  Unfortunately, multi-wavelength data cannot on its
own discriminate between these possibilities since the spectral
index for non-thermal gyrosynchotron emission can vary over a wide range
depending on the energy spectrum of the electrons involved (G\"udel 2002).
 Instead the best prospect for disentangling the relative contributions
from thermal and non-thermal free-free emission is via monitoring,
since  non-thermal emission is notably variable on timescales of months to years (Zapata et al 2004).
 
 The other factor which complicates the interpretation
of radio emission in terms of an X-ray driven wind is 
 the expected `floor' value  (equation 11)
produced by winds driven by EUV emission from $\theta_1$ C Ori. 
Comparison of equations (10) and (11) show that the floor value drops
more steeply with distance from $\theta_1$C Ori than does the X-ray wind
radio signature at fixed X-ray  luminosity. Thus  the unambiguous 
detection of X-ray driven winds is most readily achieved at relatively
large distances ( a parsec or more) from $\theta_1$C Ori.

 The detection of thermal emission of the expected magnitude in sources
more than a  parsec from $\theta_1$ C Ori would be strong evidence in favour
of X-ray driven winds interacting with the external ionising flux from
$\theta_1 $C Ori  {\footnote{From equations (7) and (10) the thermal emission from this interaction
exceeds that arising from the inner regions of the X-ray driven wind itself provided
that $L_{X30} > 10^{-3} d_{pc}^2$}.


\section{Conclusions}

  We have shown that the interaction between ionising radiation
from massive stars and  (internally driven) X-ray heated
disc winds should give rise to extended structures that
we term `X-ray proplyds'. Even in the closest suitable environment
(the ONC), resolvable X-ray proplyds would correspond only to objects
at the upper end of the X-ray luminosity function  (note that the relevant
X-ray luminosity is its value 
at a time in the past corresponding to
typical flow times from the disc to the ionisation front (hundreds
of years); this does not necessarily correspond to 
current  X-ray luminosity: see Section 3.2).
 Even in luminous X-ray sources, however, the
predicted mass loss rates are less than the rates of mass loss
driven by
external FUV heating in the close vicinity of massive stars. In the case
of the ONC, this means that the majority of proplyds (which
reside within $0.3$ pc of the central OB star $\theta_1$C Ori) are
indeed FUV driven as argued by previous authors; however 
FUV heating is ineffective at larger distances from $\theta_1$
C Ori and it is here that we expect to see a population of X-ray proplyds.

  We have demonstrated that the numbers and sizes of proplyds observed in the
ONC outside the central FUV zone is compatible with the expectations of the
X-ray proplyd model provided that we  make the
reasonable assumption that the X-ray luminosity
function was the same a few hundred years ago (when the present day
mass loss at the ionisation front was set) as it is now. Turning to 
predictions for individual objects with measured current X-ray fluxes
we find a mixed picture - in some cases ($2/8$) 
the proplyd size is a good match
to that predicted whereas in others the required X-ray flux is significantly
higher than it is now. We require greater (less) than order of magnitude
variations in $3/8$ ($3/8$) objects:  note that order of magnitude
variations in X-ray flux are commonly observed in disc bearing pre-main sequence
stars on timescales of years (Principe et al 2014).
%
We also note that at least
some of the
large proplyds in Carina (with scales of $10^3$ A.U.; Smith
et al 2003) 
observed 
up to $40$ pc from the main ionising source in this region
($\eta$  Carina) are also compatible with being
X-ray proplyds.
Future radiation-hydrodynamical modeling will 
indicate further avenues for observational characterisation of X-ray proplyds
through generation of predicted  line emission profiles and synthetic images.

 We also show that X-ray proplyds are predicted to produce thermal
radio emission at levels that are readily detectable in the
ONC. However, the predicted scaling between free-free radio luminosity
and X-ray luminosity ($L_{ff} \propto L_X ^{4/3} $ at fixed
distance from the ionising source) is similar to the trend observed
in disc-less active stars (G\"udel \& Benz 1993) and is indeed
only modestly offset towards higher radio luminosities. 
Monitoring of the emission levels on timescales of months to years
is therefore required in order to disentangle thermal and non-thermal
contributions. We also show that in order to distinguish
the signature of X-ray proplyds from that of a wind driven purely by the ionising
radiation from $\theta_1$C Ori, it is necessary to examine sources
at $> 1$ pc from $\theta_1C$ Ori. 

 Clearly we obtain the greatest model discrimination in the case that
we can detect thermal radio emission  {\it and} measure the spatial offset
of the ionisation front since in that case it is possible to 
evaluate the mass loss rate in the wind (assuming a typical  transonic
flow velocity) independent of any assumptions about the ionising flux at
the ionisation front. As already noted, the population of objects
outside the central FUV zone with optically detected offset ionisation fronts
is small (10s of objects) and corresponds (in our model) to objects with
the largest X-ray luminosities a few hundred years ago.
JWST offers the prospect of being able to image the large population
of objects with more compact structures that we predict.

\section{Acknowledgments}
We would like to acknowledge the Nordita program on Photo-Evaporation in Astrophysical Systems (June 2013) where this work was initiated and are grateful to
Will Henney for a question that sparked this investigation. 
 We gratefully acknowledge the referee, Joel Kastner, for comments
that have improved the analysis presented here and also thank Fred Adams, Richard Alexander and Ilaria Pascucci for useful
discussions. JEO 
acknowledges support through the IOA's STFC funded Visitor Programme.
This work has been supported by the DISCSIM project, grant
agreement 341137 funded by the European Research Council under
ERC-2013-ADG.}




\end{document}